# Solution of the relativistic Dirac-Mörse problem

A. D. Alhaidari

Physics Department, King Fahd University of Petroleum & Minerals, Box 5047, Dhahran 31261, Saudi Arabia

E-mail: **haidari@mailaps.org**

**Abstract:** Dirac equation for a charged particle in static electromagnetic field is written for special cases of spherically symmetric potentials. Besides the well known Dirac-Coulomb and Dirac-Oscillator potentials, we obtain a relativistic version of the S-wave Mörse potential. This is accomplished by adding a simple exponential potential term to the Dirac operator, which in the nonrelativistic limit reproduces the usual Mörse potential. The relativistic bound states spectrum and spinor wavefunctions are obtained.

The solution of Dirac-Coulomb problem including its relativistic bound state spectrum and wavefunction was established long time ago [1-3]. In 1989, the relativistic Dirac-Oscillator potential was introduced [4-8] by adding an off-diagonal linear radial term to the Dirac operator. The relativistic bound states spectrum and its eigenstates were also obtained explicitly. Taking the nonrelativistic limit reproduces the usual Schrödinger-Coulomb and Schrödinger-Oscillator solutions, respectively.

It is well established that in nonrelativistic quantum mechanics there exist exactly soluble classes of potentials (shape invariant potentials) [9] that belong to a given dynamical symmetry group. One of these classes includes the Coulomb, Oscillator, and Mörse potentials. These potentials belong to the symmetry group SO(2,1) [2] and can be transformed into one another by point canonical transformations [10]. Thus, from the point of view of group isomorphism and analyticity of parametric Lie algebras, it is tempting to look for the relativistic extension of Mörse potential. In other words, the fact that the relativistic versions of the first two problems were solved explicitly makes the solution of the third, in principle, feasible. In this letter, we succeed in finding the "Dirac-Mörse" potential and obtaining its relativistic bound states spectrum and spinor wavefunctions. This is accomplished by following the same strategy as that in the Dirac-Oscillator problem, namely by adding a radial term to the odd part of the Dirac operator, which in this case turns out to be a simple exponential. Moreover, taking the nonrelativistic limit will recover the standard Schrödinger-Mörse problem.

The general physical setting for this problem is a charged particle in static and spherically symmetric four-component electromagnetic potential. After setting up the problem, we apply a unitary transformation to Dirac equation such that the resulting second order differential equation becomes Schrödinger-like so that comparison with well-known nonrelativistic problems is transparent. Thus, simple correspondence among parameters of the two problems gives the sought after bound states spectrum and wavefunction.

**Preliminaries:** In atomic units ($m = e = \hbar = 1$) and taking the speed of light $c = \alpha^{-1}$, the Hamiltonian for a Dirac particle in four-component electromagnetic potential, $(A_0, \vec{A})$, reads

$$H = \begin{pmatrix} 1+\alpha A_0 & -i\alpha\vec{\sigma}\cdot\vec{\nabla} + \alpha\vec{\sigma}\cdot\vec{A} \\ -i\alpha\vec{\sigma}\cdot\vec{\nabla} + \alpha\vec{\sigma}\cdot\vec{A} & -1+\alpha A_0 \end{pmatrix}$$

where $\alpha$ is the fine structure constant and $\vec{\sigma}$ are the three 2×2 Pauli spin matrices. In quantum electrodynamics (the theory of interaction of charged particles with the electromagnetic field), local gauge invariance implies that the theory is invariant under the transformation

$$\left(A_0, \vec{A}\right) \to \left(A_0, \vec{A}\right) + \left(\alpha \partial \Lambda/\partial t, \vec{\nabla}\Lambda\right)$$

where $\Lambda(t,\vec{r})$ is a real space-time function. That is adding a 4-dimensional gradient of the gauge field $\Lambda(t,\vec{r})$ to the electromagnetic potential will not alter the physical content of the theory. In the lab or target frame, gauge invariance implies that the general form of the electromagnetic potential for <u>static</u> charge distribution with <u>spherical</u> symmetry is

$$\left(A_0, \vec{A}\right) = \left(\alpha V(r), \vec{0}\right) + \left(0, \vec{\nabla}\Lambda(r)\right) \equiv \left(\alpha V(r), \hat{r}W(r)\right)$$

where $\hat{r}$ is the radial unit vector. Obviously, $W(r)$ is a gauge field that does not contribute to the magnetic field. However, fixing this gauge degree of freedom by taking $W = 0$ would not have been the best choice. An alternative and proper "gauge fixing condition", which is much more fruitful, will be imposed as a constraint in equation (2) below. We will consider, however, an alternative coupling of the electromagnetic potential to the charged Dirac particle. The two off-diagonal terms $\alpha\vec{\sigma}\cdot\vec{A}$ in the Hamiltonian $H$ above are to be replaced by $\pm i\alpha\vec{\sigma}\cdot\vec{A}$, respectively, resulting in the following two-component radial Dirac equation:

$$\begin{pmatrix} 1+\alpha^2 V(r) & \alpha\left(\dfrac{\kappa}{r} + W(r) - \dfrac{d}{dr}\right) \\ \alpha\left(\dfrac{\kappa}{r} + W(r) + \dfrac{d}{dr}\right) & -1+\alpha^2 V(r) \end{pmatrix} \begin{pmatrix} g(r) \\ f(r) \end{pmatrix} = \varepsilon \begin{pmatrix} g(r) \\ f(r) \end{pmatrix} \quad (1)$$

where the spin-orbit coupling parameter $\kappa$ is defined as $\kappa = \pm (j + \tfrac{1}{2})$ for $l = j \pm \tfrac{1}{2}$ and $\varepsilon$ is the relativistic energy. This equation gives two coupled first order differential equations for the two radial spinor components. By eliminating the lower component we obtain a second order differential equation for the upper. The resulting equation may turn out not to be Schrödinger-like, i.e. it may contain first order derivatives. We apply a general local unitary transformation that eliminates the first order derivative as follows:

$$r = q(x) \quad \text{and} \quad \begin{pmatrix} g(r) \\ f(r) \end{pmatrix} = \begin{pmatrix} \cos(\rho(x)) & \sin(\rho(x)) \\ -\sin(\rho(x)) & \cos(\rho(x)) \end{pmatrix} \begin{pmatrix} \phi(x) \\ \theta(x) \end{pmatrix}$$

The stated requirement gives the following constraint:

$$\frac{dq}{dx}\left[-\alpha^2 V + \cos(2\rho) + \alpha\sin(2\rho)(W+\kappa/q) + \alpha\frac{d\rho/dx}{dq/dx} + \varepsilon\right] = \text{constant} \equiv \eta \neq 0 \quad (2)$$

This transformation and the resulting constraint are the relativistic analog of point canonical transformation in nonrelativistic quantum mechanics. Moreover, this constraint can also be thought of as the gauge fixing condition for the electromagnetic potential. In this letter, we consider the case of global unitary transformation defined by $q(x) = x$ and $d\rho/dx = 0$. Substituting this in the constraint equation (2) yields

$$W(r) = \frac{\alpha}{S} V(r) - \frac{\kappa}{r}$$

$$\eta = C + \varepsilon$$

where $S \equiv \sin(2\rho)$ and $C \equiv \cos(2\rho)$. This maps the radial Dirac equation (1) into the following:

$$\begin{pmatrix} C + 2\alpha^2 V & \alpha\left(-\frac{S}{\alpha} + \frac{\alpha C}{S} V - \frac{d}{dr}\right) \\ \alpha\left(-\frac{S}{\alpha} + \frac{\alpha C}{S} V + \frac{d}{dr}\right) & -C \end{pmatrix} \begin{pmatrix} \phi(x) \\ \theta(x) \end{pmatrix} = \varepsilon \begin{pmatrix} \phi(x) \\ \theta(x) \end{pmatrix} \quad (3)$$

which in turn gives an equation for the lower spinor component in terms of the upper:

$$\theta(r) = \frac{\alpha}{C+\varepsilon}\left[-\frac{S}{\alpha} + \frac{\alpha C}{S} V + \frac{d}{dr}\right] \phi(r) \quad (4)$$

While, the differential equation for the upper component reads

$$\left[-\frac{d^2}{dr^2} + \frac{\alpha^2}{T^2} V^2 + 2\varepsilon V - \frac{\alpha}{T} \frac{dV}{dr} - \frac{\varepsilon^2 - 1}{\alpha^2}\right] \phi(r) = 0 \quad (5)$$

where $T \equiv S/C = \tan(2\rho)$.

**The Dirac-Mörse problem:** We consider the case where the potential $V(r) = -De^{-\lambda r}$ with $D$ and $\lambda$ being real parameters. This introduces an off-diagonal exponential term in the Dirac operator as seen in equation (3). Equation (5) gives the following second order differential equation for the upper spinor component

$$\left[-\frac{d^2}{dr^2} + \left(\frac{\alpha D}{T}\right)^2 e^{-2\lambda r} - \frac{\alpha D}{T}\left(\lambda + \frac{2T}{\alpha}\varepsilon\right) e^{-\lambda r} - \frac{\varepsilon^2-1}{\alpha^2}\right] \phi(r) = 0$$

Comparing this with Schrödinger equation for the S-wave Mörse potential [9]

$$\left[-\frac{d^2}{dr^2} + B^2 e^{-2\lambda r} - B(\lambda + 2A) e^{-\lambda r} - 2E\right] \phi(r) = 0 \quad (6)$$

We obtain the following correspondence between nonrelativistic and relativistic parameters:

$$B = \alpha D/T$$
$$A = (T/\alpha)\varepsilon \quad (7)$$
$$E = (\varepsilon^2 - 1)/2\alpha^2$$

The well-known nonrelativistic bound states spectrum of equation (6) is

$$E_n = -\frac{\lambda^2}{2}\left(\frac{A}{\lambda} - n\right)^2 \quad ; n = 0, 1, 2, ..., n_{max} \leq |A/\lambda| \quad (8)$$

The substitution (7) gives the following relativistic spectrum

$$\varepsilon_n = \frac{1}{1+T^2}\left[\alpha \lambda T n + \sqrt{1 + T^2 - (\alpha \lambda n)^2}\right]$$

where $n = 0, 1, 2, ..., n_{max} \leq |1/\alpha\lambda|\sqrt{1+T^2}$. Taking the nonrelativistic limit of this spectrum with

$$\alpha \to 0$$
$$\varepsilon_n \approx 1 + \alpha^2 E_n$$
$$T \approx \alpha\tau$$

reproduces the nonrelativistic spectrum (8) with $\tau = A$. The bound states wavefunction of the nonrelativistic problem [9] is mapped, using (7), into the following upper spinor component wavefunction

$$\phi_n(r) = a_n(\mu e^{-\lambda r})^{v_n/2} \exp(-\tfrac{1}{2}\mu e^{-\lambda r}) L_n^{v_n}(\mu e^{-\lambda r})$$

where $L_n^v(x)$ is the generalized Laguerre polynomial [11], $a_n$ is the normalization constant, and

$$\mu = 2(\alpha/\lambda T)D$$
$$v_n = 2\left(\frac{T}{\alpha\lambda}\varepsilon_n - n\right)$$

Equation (4) gives the lower spinor component in terms of the upper as

$$\theta_n(r) = \frac{\alpha}{\varepsilon_n + C}\left(-\frac{S}{\alpha} - \frac{\alpha}{T}De^{-\lambda r} + \frac{d}{dr}\right)\phi_n(r)$$

Using the differential and recursion properties of the Laguerre polynomials [11], we can write it explicitly as

$$\theta_n(r) = -Ta_n(\mu e^{-\lambda r})^{v_n/2} \exp(-\tfrac{1}{2}\mu e^{-\lambda r})\left[L_n^{v_n}(\mu e^{-\lambda r}) + \frac{(\alpha\lambda/T)n - 2\varepsilon_n}{C + \varepsilon_n} L_{n-1}^{v_n}(\mu e^{-\lambda r})\right]$$

It is instructive, at this point, to apply the above technique and reproduce the solutions for the other two problems in this class of relativistic potentials using the same notation. The potential functions $V(r)$ and $W(r)$, physical parameters and bound states spectrum obtained for all three problems are listed in the table.

**The Dirac-Coulomb problem:**
The 2$^{nd}$ order differential equation for this problem and for a general angular momentum, is as follows [1-3]:

$$\left[-\frac{d^2}{dr^2} + \frac{\gamma(\gamma+1)}{r^2} + 2\frac{Z\varepsilon}{r} - \frac{\varepsilon^2 - 1}{\alpha^2}\right]\phi(r) = 0$$

where $Z$ is the particle charge number and $\gamma$ is the relativistic angular momentum which is related to the spin-orbit coupling parameter $\kappa$ via the relation $\kappa^2 = \gamma^2 + \alpha^2 Z^2$. The two components of the radial spinor wavefunction are:

$$\phi_n(r) = a_n(\lambda_n r)^{\gamma+1} e^{-\lambda_n r/2} L_n^{2\gamma+1}(\lambda_n r)$$

$$\theta_n(r) = \frac{\alpha\lambda_n/2}{\varepsilon_n + \gamma/\kappa} a_n(\lambda_n r)^{\gamma} e^{-\lambda_n r/2} \times$$
$$\left[(1 - 2Z/\kappa\lambda_n)(2\gamma + n + 1)L_n^{2\gamma}(\lambda_n r) + (1 + 2Z/\kappa\lambda_n)(n+1)L_{n+1}^{2\gamma}(\lambda r)\right]$$

where

$$\lambda_n = -\frac{2Z\varepsilon_n}{\gamma + n + 1}$$

**The Dirac-Oscillator problem:**
The 2$^{nd}$ order differential equation for this problem and again for a general angular momentum, is as follows [4-8]:

$$\left[-\frac{d^2}{dr^2}+\frac{\kappa(\kappa+1)}{r^2}+\zeta^2 r^2+(2\kappa-1)\zeta-\frac{\varepsilon^2-1}{\alpha^2}\right]\phi(r)=0$$

where $\zeta$ is the oscillator strength parameter and $\kappa$ is the spin-orbit coupling parameter defined as $\kappa = \pm (j + \tfrac{1}{2})$ for $l = j \pm \tfrac{1}{2}$. The two components of the radial spinor wavefunction are:

$$\phi_n(r) = a_n(\sqrt{\zeta}\,r)^{\kappa+1} e^{-\zeta r^2/2} L_n^{\kappa+1/2}(\zeta r^2)$$

$$\theta_n(r) = \frac{2\alpha\sqrt{\zeta}}{\varepsilon_n+1}(n+\kappa+1/2) a_n (\sqrt{\zeta}\,r)^{\kappa} e^{-\zeta r^2/2} L_n^{\kappa-1/2}(\zeta r^2)$$

Finally, we like to make three comments which are relevant for fruitful extension of the present work. The first, is that supersymmetric quantum mechanics should be a suitable setting for extending this work to investigate other classes of relativistic potentials; for example, "Dirac-Pöschl-Teller" and "Dirac-Rosen-Mörse", etc. Secondly, an interesting problem is to account for all relativistic potentials that belong to a given class and find the corresponding symmetry group and/or its relativistic extension. Thirdly, relativistic scattering can also be set up where a general spin-dependent perturbing short-range potential is superimposed on the reference Dirac Hamiltonian, $H_0$, which includes one of these relativistic potentials. Such program has already been developed for the Dirac-Coulomb problem using the relativistic J-matrix method of scattering [12]. The same can, in principle, be applied to Dirac-Mörse potential scattering. In this formalism a proper $L^2$ spinor basis is chosen such that the matrix representation of the reference Hamiltonian is tridiagonal rendering the $H_0$-problem analytically soluble. The solutions of the resulting three-term recursion relation give the J-matrix kinematical coefficients necessary for scattering calculation.

**Acknowledgements:** The author is indebted to Dr. H. A. Yamani for very enlightening discussions and improvements on the original manuscript.

**Table Caption:**

The potential functions *V(r)* and *W(r)*, physical parameters, and bound states spectrum obtained using the technique developed in this letter for all three problems. In the table:

$$\kappa^2 = \gamma^2 + (\alpha Z)^2$$
$$\beta^2 = B^2 + (\alpha D)^2$$

**Table**

|  | **Dirac-Coulomb** | **Dirac-Oscillator** | **Dirac-Mörse** |
|---|---|---|---|
| $V(r)$ | $Z/r$ | $0$ | $-De^{-\lambda r}$ |
| $W(r)$ | $0$ | $\zeta r$ | $-\beta e^{-\lambda r} - \kappa/r$ |
| $S$ | $\alpha Z/\kappa$ | $0$ | $\alpha D/\beta$ |
| $C$ | $\gamma/\kappa$ | $1$ | $B/\beta$ |
| $\varepsilon_n$ | $\left[1+\left(\dfrac{\alpha Z}{\gamma+n+1}\right)^2\right]^{-1/2}$ | $\sqrt{1+2\alpha^2\zeta(2n+l+\kappa+1)}$ | $\dfrac{B}{\beta^2}\left[\alpha^2\lambda Dn+\sqrt{\beta^2-(\alpha\lambda Bn)^2}\right]$ |